\documentclass[11pt]{article}

\usepackage{cite}
 \usepackage{subfigure}
\usepackage{multirow}
\usepackage{helvet}
\usepackage{amsmath}
\usepackage{amssymb}
\usepackage{setspace}
\usepackage{graphicx}
\usepackage{empheq}
\usepackage{soul}
\usepackage{tabularx}
\usepackage{array}
\usepackage{float}
\usepackage{braket}
\usepackage[utf8]{inputenc}
\usepackage{url}
\usepackage{hyperref}
\usepackage{slashed}
\usepackage[font=footnotesize,labelfont=bf]{caption}
\usepackage{color}

\def\be{\begin{equation}}
\def\ee{\end{equation}}

\usepackage[a4paper,top=3cm,bottom=3cm,left=2.5cm,right=2.5cm,bindingoffset=0mm]{geometry}

\begin{document}

\begin{center}
{\Large \bf Higher precision constraints on the tau $g-2$  in \\ \vspace*{0.1cm} LHC photon--initiated production: \\ \vspace*{0.2cm} a full account of hadron dissociation and soft survival effects}

\vspace*{1cm}

{\sc 
L.A.~Harland-Lang%
\footnote[1]{email: l.harland-lang@ucl.ac.uk}
}
\vspace*{0.5cm}

{\small\sl
  Department of Physics and Astronomy, University College London, London, WC1E 6BT, UK

}

\begin{abstract}
\noindent We present  the first calculation of photon--initiated $\tau$ pair production in the presence of non--zero anomalous magnetic ($a_\tau$) and/or electric dipole ($d_\tau$) moments of the $\tau$ lepton that accounts for the non--trivial interplay between these modifications with the soft survival factor and the possibility of dissociation of the hadron (proton or ion) beam. The impact of these is on general grounds not expected to have a uniform dependence on the value of $a_\tau, d_\tau$, but in all previous analyses this assumption has been made.  We have therefore investigated the importance of these effects in the context of photon--initiated $\tau$ pair production in both pp and PbPb collisions. This is in general found to be relatively small, at the percent level in terms of any extracted limits or observations of $a_\tau, d_\tau$, such that these effects can indeed be safely ignored in existing experimental analyses. However, as the precision of such determinations increases in the future, the relevance of these effects will likewise increase. With this in mind we have made our calculation publicly available in the \texttt{SuperChic} Monte Carlo generator, including the possibility to simulate this process for varying $a_\tau, d_\tau$ without rerunning. 

\end{abstract}

\end{center}

\section{Introduction}

The LHC is a collider of electromagnetically charged proton and heavy ions and as such, as well as being a QCD machine, it can effectively act as a photon--photon collider. This photon--initiated (PI) particle production provides a unique probe of physics within and beyond the SM, see e.g.~\cite{Harland-Lang:2020veo} for further discussion and references, and~\cite{LHCForwardPhysicsWorkingGroup:2016ote,Bruce:2018yzs,Vysotsky:2018slo,Schoeffel:2020svx} for reviews. A key feature of these processes is the colour--singlet nature of the photon, which allows for PI production to occur in association with no further colour flow between the colliding hadrons. As a result, for heavy ion beams PI production can lead to extremely clean ultraperipheral collisions (UPCs), where no other particles other than the produced state in the photon--photon collision is present. For proton beams, PI production can occur both exclusively or semi--exclusively, i.e. with or without the proton remaining intact, respectively, and hence PI production can be selected by tagging the intact protons~\cite{AFP,Albrow:1753795} and/or selecting for events with no additional associated charged tracks in the central detector.

A topical example is the PI production of $\tau$ lepton pairs, $\gamma\gamma \to \tau^+\tau^-$. The motivation for measuring this process in PI production was  discussed in e.g.~\cite{delAguila:1991rm,Atag:2010ja} and more recently in~\cite{Beresford:2019gww} (see also~\cite{Dyndal:2020yen}). In these recent studies,  the potential for a measurement of this process to significantly improve on the at the time best constraints on the $\tau$ anomalous magnetic moment, $a_\tau$, from DELPHI~\cite{DELPHI:2003nah} at LEP, as well as the potential improved sensitivity to the $\tau$ electric dipole moment, $d_\tau$, was demonstrated. The anomalous magnetic moment of the $\tau$ is in particular much less well constrained experimentally than in the electron and muon cases, which are now measured to a precision of twelve~\cite{Parker:2018vye,Fan:2022eto} and ten~\cite{Muong-2:2021ojo,Muong-2:2023cdq,Aoyama:2020ynm} significant digits, respectively. For the $\tau$, on the other hand, its short lifetime precludes the use of equivalent storage ring probes, and hence the  best pre--LHC experimental limit~\cite{DELPHI:2003nah}  is roughly an order of magnitude higher than the SM predicted value~\cite{Eidelman:2007sb} of $a_\tau=1.17721\pm 0.00005 \times 10^{-3}$. It is therefore 
very well motivated to improve on these constraints, given in particular the possibility for BSM effects to scale with the lepton mass, and hence be enhanced in the $\tau$ sector. The electric dipole moment, $d_\tau$, is on the other hand predicted to be highly suppressed in the SM~\cite{Yamaguchi:2020eub}, and so again improving on current constraints for this can provide a probe of BSM physics. 

ATLAS~\cite{ATLAS:2022ryk} and CMS~\cite{CMS:2022arf} have presented measurements of $\tau$ lepton pair production in UPCs, with the ATLAS constraints being comparable to those at LEP. More recently,  PI $\tau$ pair production has been observed for the first time in $pp$ collisions by CMS~\cite{CMS:2024skm} (see also~\cite{Beresford:2024dsc} for recent discussion). This measurement significantly improves on previous constraints on the $\tau$ anomalous magnetic moment, and provides constraints on it at a level that is close to the SM prediction. 

The above analyses, however, rely on certain simplifying assumptions about the signal modelling as both $a_\tau$ and $d_\tau$ are varied. One aspect of this relates to the `survival factor' probability that the colliding hadrons do not interact strongly, leading to colour flow between the hadrons and a high multiplicity event that will fail the  selection applied to isolate PI production. As discussed in detail in~\cite{Harland-Lang:2018iur,Harland-Lang:2020veo,Harland-Lang:2021ysd} and references therein, this is not a constant probability but rather depends sensitively on the produced particles, the final--state kinematics and whether the outgoing hadrons dissociate or not. The process dependence is in particular omitted in the treatments of~\cite{Dyndal:2020yen} and~\cite{Shao:2022cly}, which are used in the ATLAS PbPb~\cite{ATLAS:2022ryk} and CMS pp~\cite{CMS:2024skm} determinations, respectively.
 This is particularly relevant, as in general the survival factor, and its kinematic dependence, will vary with $a_\tau, d_\tau$, and hence this omission may effect the corresponding limits, or any future measurement, of these. 
 
 In addition, in the CMS pp analysis~\cite{CMS:2024skm}, events are selected by applying a veto on additional charged tracks (i.e. without tagging protons), and hence as well as purely elastic (EL) production, with both protons remaining intact, single and double dissociative (SD and DD) production, where one or both protons dissociate, will contribute to the signal. Indeed, as noted in~\cite{Harland-Lang:2020veo,Bailey:2022wqy}, the dissociative contribution under these conditions is in particular generally larger than the purely elastic. Indeed, this is indirectly observed in~\cite{CMS:2024skm} in the case of muon pair production, where the ratio of the observed signal to the predicted elastic component is of order three or larger, depending on the dimuon mass. In this CMS analysis, the corresponding $\tau$ pair production cross section is reweighted by this measured ratio in order to convert the data back to a purely elastic component. However, this again assumes that the relative contribution from dissociative production is independent of the values of $a_\tau, d_\tau$. More precisely, recalling that the muon anomalous magnetic moment is strongly suppressed with respect to the SM prediction for the $\tau$, these are effectively set to zero in this conversion.
 
 Finally, in~\cite{Harland-Lang:2023ohq} a full treatment of UPCs including mutual ion dissociation was presented. In the current context, it may in principle be the case that the relative event fractions with or without neutron emission, due to ion dissociation, are also dependent on  $a_\tau, d_\tau$. We also investigate this possibility here.
 
In this paper we will present a complete account of the above effects. As we will see, their impact is generally small, and  within the other uncertainties on the current limits set in pp and PbPb collisions. However, in the future, as such limits or indeed any eventual observation become increasingly precise, this may not be the case. With this in mind, we provide a full implementation of the $a_\tau, d_\tau$ dependent cross sections in UPCs, and in pp collisions with and without proton dissociation, in the publicly available \texttt{SuperChic} Monte Carlo (MC) generator~\cite{SuperCHIC}.

The outline of this paper is as follows. In Section~\ref{sec:theory} we describe the underlying theory behind the modelling of photon--initiated $\tau$ pair production in the presence of anomalous magnetic and electric dipole moments. In Section~\ref{sec:res} we present the results of the \texttt{SuperChic}  implementation of this. Finally, in Section~\ref{sec:conc} we conclude.

\section{Theory}\label{sec:theory}

\subsection{The $\tau$ pair production amplitude}

The $\tau$ anomalous magnetic and dipole moments enter the QED Lagrangian via
\be
  \mathcal{L} = \tfrac{1}{2} \bar{\tau}_\text{L}\sigma^{\mu\nu}  \left(a_\tau \tfrac{e}{2m_\tau} - \mathrm{i} d_\tau \gamma_5 \right) \tau_\text{R} F_{\mu\nu}.
\ee
where $\tau_\text{L,R}$ are left and right handed tau spinors and $\sigma^{\mu\nu} = i[\gamma^\mu, \gamma^\nu]/2 $. We follow the approach of~\cite{Beresford:2019gww} to introduce BSM modifications of $\delta a_\tau$ and  $\delta d_\tau$, namely via SM effective field theory (SMEFT)~\cite{Escribano:1993pq}. The corresponding BSM Lagrangian consists of a dimension--6 operator that modifies $a_\tau$ and $d_\tau$ at tree level
\begin{align}
    \mathcal{L}_\text{SMEFT} = \left(C_{\tau B} /\Lambda^2\right) \bar{L}_{\tau}\sigma^{\mu\nu} \tau_R H B_{\mu\nu}
    \label{eq:BSMLagrangian}
\end{align}
where $L_\tau$ ($H$) is the tau-lepton (Higgs) doublet, $B_{\mu\nu}$ is the hypercharge field, and $C_{\tau B}$ is the complex Wilson coefficient in the Warsaw basis~\cite{Grzadkowski:2010es}. The real and imaginary  parts of $C_{\tau B}$ correspond to the shifts
\begin{align}
    \delta a_\tau &= \frac{2m_\tau}{e}\frac{\mathrm{Re}\left[C_{\tau B}\right]}{M},\quad
    \delta d_\tau = \frac{\mathrm{Im}\left[C_{\tau B}\right]}{M},
    \label{eq:delta_a_d_tau_defn}
\end{align}
such that the corresponding $ \tau \overline{\tau}\gamma$ vertex has the form
\be\label{eq:vtau}
V_{\tau\tau \gamma}^\mu= i e\gamma^\mu - \left[\delta a_\tau\frac{e}{2 m_\tau}+i \delta d_\tau\gamma_5\right]\sigma^{\mu\nu}q_\nu\;,
\ee
where $q$ is the photon momentum and we have also included the usual LO contribution in the SM. The corresponding structure of the $\gamma\gamma\to \tau^+ \tau^-$ amplitude then has the form
\be\label{eq:mtautau}
\mathcal{M}_{\mu\nu}= \mathcal{M}^{0,0}_{\mu\nu}+a_\tau \mathcal{M}^{1,0}_{\mu\nu}+d_\tau \mathcal{M}^{0,1}_{\mu\nu}+a_\tau d_\tau \mathcal{M}^{1,1}_{\mu\nu}+a_\tau^2 \mathcal{M}^{2,0}_{\mu\nu}+d_\tau^2 \mathcal{M}^{0,2}_{\mu\nu}\;,
\ee
and the cross section in general receives contribution up to $O(a_\tau^4, d_\tau^4)$. The above formalism is implemented, as proposed in~\cite{Beresford:2019gww} using the \texttt{SMEFTsim\_general\_alphaScheme\_UFO} model of the \textsc{SMEFTsim} package~\cite{Brivio:2017btx,Brivio:2020onw}, in  \textsc{MadGraph\_5\_}a\textsc{mc@NLO}~\cite{Alwall:2011uj,Alwall:2014hca}. The latter output is, as in~\cite{Bailey:2022wqy}, evaluated in standalone mode, such that the resulting amplitudes can be directly interfaced to \texttt{SuperCHIC}. For simplicity, in the results which follow we will denote the $\delta a_\tau, \delta d_\tau$ that enter the above expressions as simply $a_\tau, d_\tau$.

\subsection{Modelling photon--initiated production}

To model photon--initiated $\tau$ pair production we apply the approach described in~\cite{Harland-Lang:2019eai,Harland-Lang:2021ysd}. Here we very briefly summarise the key ingredients of this, but refer the reader to these references for further details. The key formula for the PI cross section of $\tau$ pairs in the high energy limit is given by 
  \be\label{eq:sighh}
  \sigma_{pp} = \frac{1}{2s} \int \frac{{\rm d}^3 p_1' {\rm d}^3 p_2' {\rm d}\Gamma}{E_1' E_2'}   \alpha(Q_1^2)\alpha(Q_2^2)
  \frac{\rho_1^{\mu\mu'}\rho_2^{\nu\nu'} M^*_{\mu'\nu'}M_{\mu\nu}}{Q_1^2 Q_2^2}\delta^{(4)}(q_1+q_2 - k)\;,
 \ee
 where the outgoing hadronic systems have momenta $p_{1,2}'$ and the photons have momenta $q_{1,2}$, with $q_{1,2}^2 = -Q_{1,2}^2$. We consider the production of a system of 4--momentum $k =  k_1 +k_2$ where ${\rm d}\Gamma = {\rm d}^3 k_1 {\rm d}^3 k_2 / [4 E_1 E_2 (2\pi)^6]$ is the standard two--body phase space volume for the production of $\tau$ leptons with momenta $k_{1,2}$. $M^{\mu\nu}$ corresponds to the $\gamma\gamma \to \tau^+ \tau^-$ production amplitude, with arbitrary photon virtualities, which is given as in \eqref{eq:mtautau}.
 
  In the above expression, $\rho$ is the density matrix of the virtual photon, which is given by
 \be\label{eq:rho}
 \rho_i^{\alpha\beta}=2\int \frac{{\rm d}M_i^2}{Q_i^2}  \bigg[-\left(g^{\alpha\beta}+\frac{q_i^\alpha q_i^\beta}{Q_i^2}\right) F_1(x_{B,i},Q_i^2)+ \frac{(2p_i^\alpha-\frac{q_i^\alpha}{x_{B,i}})(2p_i^\beta-\frac{q_i^\beta}{x_{B,i}})}{Q_i^2}\frac{ x_{B,i} }{2}F_2(x_{B,i},Q_i^2)\bigg]\;,
 \ee
where $x_{B,i} = Q^2_i/(Q_i^2 + M_{i}^2 - m_p^2)$ for a hadronic system of mass $M_i$ and we note that the definition of the photon momentum $q_i$ as outgoing from the hadronic vertex is opposite to the usual DIS convention. Here, the integral over $M_i^2$ is understood as being performed simultaneously with the phase space integral over $p_{i}'$, i.e. is not fully factorized from it (the energy $E_i'$ in particular depends on $M_i$). 

By suitably substituting for the relevant elastic and/or inelastic proton structure functions $F_{1,2}$ in the above expression, as described in detail in~\cite{Harland-Lang:2019eai}, we can then provide predictions for elastic (EL), single (SD) and double dissociative (DD) production. For the case of heavy ion collisions, on the other hand, we are at this stage only interested in elastic production and we have 
 \be
 F_2(x_{B,i},Q_i^2) = F_p^2(Q_i^2)G_E^2(Q_i^2)\, \delta(1-x_{B,i})\;,
 \ee
where $F_p^2(Q^2)$ is the squared form factor of the ion, which is given in terms of the proton density in the ion, $\rho_p(r)$, and $G_E$ is the  `Sachs' form factor of the of the protons within the ion, see~\cite{Harland-Lang:2021ysd} for details.

To account for the survival factor, that is the probability of no additional inelastic hadron--hadron interactions, we work at the level of the $\gamma\gamma \to \tau^+ \tau^-$ production amplitude. Following the discussion in~\cite{Bailey:2022wqy}, we can in particular write the dominant contribution to the cross section as 
\be\label{eq:sigpps2a}
 \sigma_{pp} = \frac{1}{8 \pi^2 s}  \int  {\rm d}x_1 {\rm d}x_2\,{\rm d}^2 q_{1_\perp}{\rm d}^2 q_{2_\perp
} {\rm d \Gamma}  \frac{{\rm d}M_1^2}{Q_1^2}  \, \frac{{\rm d}M_2^2}{Q_2^2} \, \frac{1}{\tilde{\beta}} \,|T(q_{1_\perp},q_{2_\perp})|^2 \delta^{(4)}(q_1+q_2 - p_X)\;,
 \ee
 where
  \be
 x_{1,2} = \frac{1}{\sqrt{s}}\left(E_{X} \pm p_{X,z}\right) = \frac{m_{X_\perp}}{\sqrt{s}} e^{\pm y_{X}}\;,
 \ee
 with $X=\tau^+\tau^-$ and $q_{i\perp}$ are the photon transverse momenta, while $\tilde{\beta}$ is defined in~\cite{Harland-Lang:2019zur}. Here we have 
 \be\label{eq:tqq}
T(q_{1_\perp},q_{2_\perp})\propto \frac{q_{1_\perp}^\mu q_{2_\perp}^\nu}{Q_1^2 Q_2^2} M_{\mu\nu}\;,
 \ee 
where we omit the full set of kinematic arguments of $T$ for brevity, see~\cite{Bailey:2022wqy} for further details. To account for the survival factor we then simply replace our expression for $T$ with one that accounts for the `rescattering' effects of potential hadron--hadron interactions
\be
T(q_{1_\perp},q_{2_\perp})\to T(q_{1_\perp},q_{2_\perp})+T^{\rm res}(q_{1_\perp},q_{2_\perp})\;,
\ee
where $T^{\rm res}$ is given in terms of the original amplitude $T$ and the elastic hadron--hadron scattering amplitude, $T_{\rm el}$, via
\begin{equation}\label{skt}
T^{\rm res}({q}_{1_\perp},{q}_{2_\perp}) = \frac{i}{s} \int\frac{{\rm d}^2  {k}_\perp}{8\pi^2} \;T_{\rm el}(k_\perp^2) \;T({q}_{1_\perp}',{q}_{2_\perp}')\;,
\end{equation}
where $q_{1_\perp}=q_{1_\perp}'+k_\perp$ and $q_{2_\perp}'=q_{1_\perp}-k_\perp$. The elastic amplitude is given in terms of the Fourier transform of the hadron--hadron opacity,  $\exp(-\Omega_{hh}(s,b_\perp))$, which  represents the probability that no inelastic scattering occurs at impact parameter $b_\perp$. While the above discussion relates to the dominant part of the PI amplitude, in the actual calculation the full amplitude is accounted for following the approach described in~\cite{Bailey:2022wqy}. In addition, as discussed in detail in~\cite{Harland-Lang:2023ohq}, for heavy ion collisions this can be suitably modified to also include the probability that the ions do or do not undergo mutual dissociation, due to addition photon exchanges, resulting in a certain number of additional neutron emissions.

From the discussion above, we can see that in proton--proton collisions the relative cross sections for EL, SD and DD production will depend on the precise form of the $\gamma\gamma \to \tau^+\tau^-$ amplitude, as it enters in \eqref{eq:sighh}. In particular, we can see from \eqref{eq:vtau} that each additional factor of $\delta a_\tau, \delta d_\tau$ introduces an additional factor of the photon momentum $q$ in the numerator. Given the inelastic proton structure functions fall rather less steeply with the photon $Q_i^2$ than in the elastic case, we may therefore expect these contributions to be somewhat enhanced in the SD and DD channels.

In terms of the impact of the survival factor, we can see from \eqref{skt} that this will depend on the form of the amplitude $T$, which is given in terms of the $\gamma\gamma \to \tau^+\tau^-$  amplitude via \eqref{eq:tqq}. Given the differing Lorentz structure of the terms in \eqref{eq:vtau} that are $\sim \delta a_\tau, \delta d_\tau$ in comparison to the LO SM contribution, we may again expect the impact of survival effects to be different for these.

\section{Results}\label{sec:res}

In this section we show a selection of results for $\tau$ pair production in both pp and PbPb collisions, in all cases implemented in  \texttt{SuperCHIC}  following the approach described in the previous sections. To focus most directly on the impact of proton dissociation and survival effects on the production cross section in the presence of anomalous magnetic and electric dipole moments, we will define our fiducial cross sections and show our results at the level of $\tau$--level pseudo-observables, prior to their decay. However, we note that the implementation in \texttt{SuperCHIC} provides full final--state particle information in the \texttt{LHE}~\cite{Alwall:2006yp} and \texttt{HepMC}~\cite{Dobbs:2001ck}  unweighted event formats, such that the $\tau$ decays can be fully accounted for in any comparison to data. Moreover, the visible mass and lepton $p_\perp$ from the $\tau$ decays can act as (biased) proxies for the $\tau$ pair invariant mass and $\tau$ lepton $p_\perp$ pseudo--observables that we will consider below.

Focussing on the case of non--zero $a_\tau$, i.e. assuming $d_\tau=0$ for now, we will in all cases make use of the fact that following \eqref{eq:mtautau} the cross section can be written in the form
\be\label{eq:sigtau}
{\rm d} \sigma = \sum_{i=0}^4  (a_\tau)^i {\rm d} \sigma_i\;,
\ee
where  we denote the cross section as ${\rm d} \sigma$ to indicate that this applies differentially, i.e. for a binned observable.  Each individual contribution, ${\rm d} \sigma_i$, is gauge invariant and can be individually extracted from the simulation by isolating the contributing squared amplitudes and interference terms that contribute. We apply this approach, and make this possibility available in the \texttt{SuperCHIC} for the user. Once this is done, the combined cross section, ${\rm d}\sigma$, can be evaluated for arbitrary values of $a_\tau$ without recalculating the results, while the individual contributions can also be considered in isolation in order to clarify the analysis. For the case of $d_\tau$  the expansion is identical to \eqref{eq:sigtau} but the linear and cubic terms in $d_\tau$ are zero due to the presence of the $\gamma_5$ in \eqref{eq:vtau}. The non--zero coefficients, ${\rm d} \sigma_i$ are identical after accounting for the appropriate rescaling of the normalization as derived from \eqref{eq:vtau}. 

\begin{figure}
\begin{center}
\includegraphics[scale=0.62]{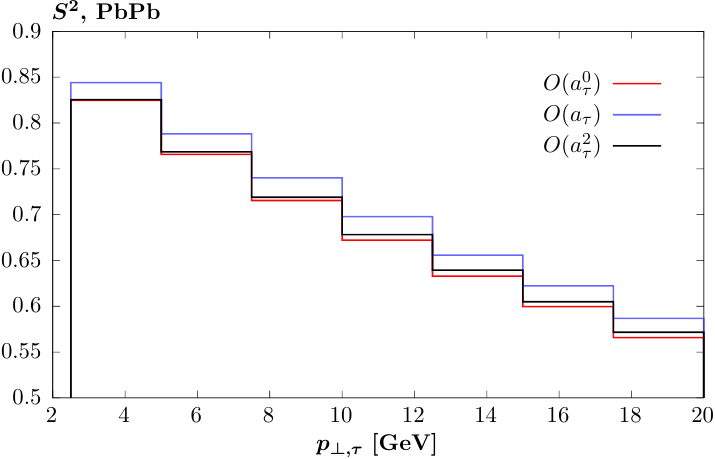}
\caption{\sf \footnotesize{Survival factor for $\tau$ pair production in PbPb collisions at $\sqrt{s_{NN}}=5.02$ TeV, for different $O(a_\tau)$ contributions to the overall cross section. Cuts are as described in text.}}
\label{fig:PbPb_ai_surv}
\end{center}
\end{figure}

We begin by considering PbPb collisions, at $\sqrt{s_{NN}}=5.02$ TeV, for $|\eta_\tau|<2.5$ and $p_\perp^\tau < 2.5$ GeV.  In Fig.~\ref{fig:PbPb_ai_surv} we show the survival factor differentially in the $\tau$ transverse momentum, $p_{\perp,\tau}$, for the individual components of \eqref{eq:sigtau}. We do not show the $O(a_\tau^3,a_\tau^4)$ cases as these are generally subleading, and give a very similar result to the $O(a_\tau^2)$ component for the survival factor. Any difference between the $O(a_\tau)$ and higher components and the $O(a_\tau^0)$ case will be an effect that is missed in e.g. the calculation of~\cite{Dyndal:2020yen} that has been applied in the ATLAS  analyses~\cite{ATLAS:2022ryk}, as described in the introduction.

We can see that this is indeed not identical between the three considered cases, with the $O(a_\tau)$ survival factor being $\sim 5\%$ higher. This is as expected given the form of the vertex \eqref{eq:vtau}, and in particular the presence of the photon 4--momentum $q^\mu$. In impact parameter space, this will lead to a factor of $b_\perp^\mu$, such that the amplitude vanishes at zero impact parameter where the impact of survival effects is larger, see~\cite{Kaidalov:2003fw} for early discussion of this effect. The $O(a_\tau^2)$ component is on the other hand rather close to the $O(a_\tau^0)$ one, albeit slightly higher than it at higher $p_{\perp,\tau}$. This is of particular relevance as it is in fact this  $O(a_\tau^2)$ component that provides by far the dominant deviation due to a non--zero $a_\tau$ at higher $p_{\perp,\tau}$. We may therefore expect the impact of this on any $a_\tau$ determination to be relatively small. 

To examine if this is the case  we will consider cross section modifications, defined as
\be
\delta = \frac{1}{{\rm d} \sigma_0}\sum_{i=1}^4  (a_\tau)^i {\rm d} \sigma_i\;,
\ee
that is, these define the relative deviation with respect to the $a_\tau=0$ cross section that comes from a particular value of $a_\tau$. We can then examine the impact that including a full modelling of e.g. survival effects will have on this deviation by taking the ratio of the $\delta$ with this accounted for to the case without, for a given of value of $a_\tau$. In particular, while in e.g. the calculation of~\cite{Dyndal:2020yen} an evaluation of the survival factor is included, its process (and in particular $a_\tau$) dependence is not, and hence the relative impact on the differing terms in \eqref{eq:sigtau} will be constant. The no survival factor case, where this will also by definition be true, therefore effectively corresponds to this case for the value of $\delta$ defined above. We note that while the above analysis provides a good estimate for the difference in differential distribution due to the inclusion of the process dependence of the survival factor on the non--zero $O(a_\tau)$ and above components, it is is also in general true that there will differences with respect to the $O(a_\tau^0)$ term, i.e. the LO SM prediction. We do not consider a comparison of this effect here.

\begin{figure}
\begin{center}
\includegraphics[scale=0.62]{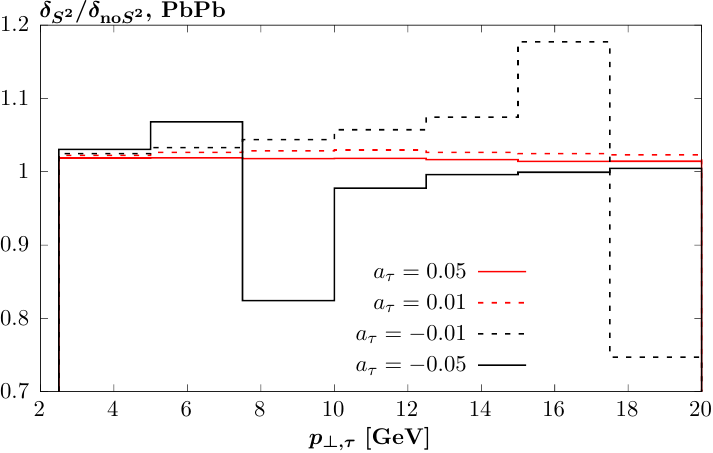}
\includegraphics[scale=0.62]{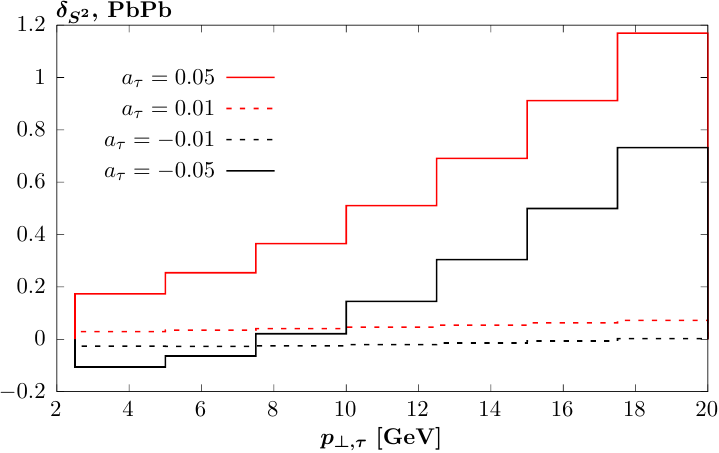}
\caption{\sf \footnotesize{Difference in cross section modifications, $\delta$, due to non--zero anomalous $a_\tau$, between the case with and without the survival factor included. Results shown for  $\tau$ pair production in PbPb collisions at $\sqrt{s_{NN}}=5.02$ TeV, as in Fig.~\ref{fig:PbPb_ai_surv}}}
\label{fig:PbPb_ai_del}
\end{center}
\end{figure}

This ratio is shown in Fig.~\ref{fig:PbPb_ai_del} (left) and we  can see that it is indeed often true that this modification is at the percent level. The larger differences observed for negative values of $a_\tau$ are due to the fact that the  value of $\delta$ passes through zero in this region, as is seen in Fig.~\ref{fig:PbPb_ai_del} (right). While it is not straightforward to immediately translate these results into a corresponding change in a given determination of  (or as is currently the case experimentally, limit setting on) $a_\tau$, we can make a few general comments here. First, we note that if the ratio plotted in Fig.~\ref{fig:PbPb_ai_del} is roughly constant with $p_{\perp,\tau}$ then this does not correspond to a constant offset in between the resulting differential cross sections. Rather, for $a_\tau=0.01$ and 0.05, as the ratio is constant and larger than unity, from the right plot we can see that these distributions will fall somewhat more slowly with $p_{\perp,\tau}$. Under the (incorrect) assumption that the $O(a_\tau)$ component provides the dominant correction, then the corresponding fit to $a_\tau$, or its limit, would be shifted down by the ratio shown in the left plot. In reality, given the presence of the higher power term in $a_\tau$, this correction will not be so direct, and more generally in the presence of a non--constant offset the impact will again be correspondingly non--trivial. 

More generally, the precise impact will depend on the specific experimental binning, whether the shape information or also the absolute value of the cross section is used, and most significantly the corrections due to the decay of the $\tau$ leptons and construction of the corresponding experimental observables due to this. Given these factors, and the relatively mild impact of the effects observed in Fig.~\ref{fig:PbPb_ai_del} and below, we do not pursue a full evaluation of the impact on any $a_\tau$ determination here. However, as a guide we note that following the discussion above a percent level deviation in the ratios in the left plot will roughly speaking correspond to a percent level deviation in the value of $a_\tau$, or the limit on it, that is extracted.

\begin{figure}
\begin{center}
\includegraphics[scale=0.62]{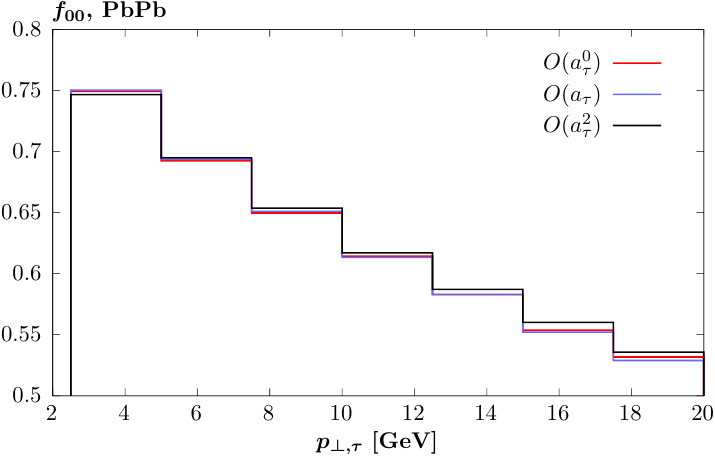}
\includegraphics[scale=0.62]{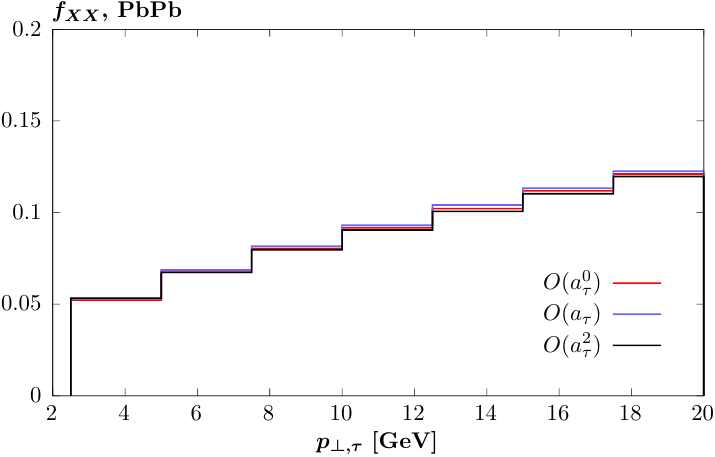}
\caption{\sf \footnotesize{Ratios of $00$ and $XX$ to the total cross sections for $\tau$ pair production in PbPb collisions at $\sqrt{s_{NN}}=5.02$ TeV, for different $O(a_\tau)$ contributions to the overall cross sections. Cuts are as described in text.}}
\label{fig:PbPb_ai_XX}
\end{center}
\end{figure}

Next, in Fig.~\ref{fig:PbPb_ai_XX} we show the same cross section components as in Fig.~\ref{fig:PbPb_ai_surv}, but now look at the predicted $0n0n$ and $XnXn$ fractions (see~\cite{Harland-Lang:2023ohq}). We we can see that these are in fact expected to be rather uniform between the different ${\rm d} \sigma_i$ components, and hence this effect is not expected to play a noticeable role in the determination of $a_\tau$. We therefore do not consider it further in what follows.

\begin{figure}
\begin{center}
\includegraphics[scale=0.62]{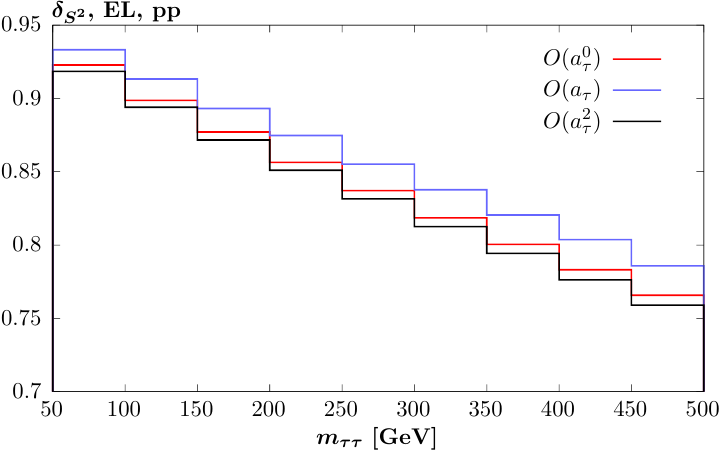}
\includegraphics[scale=0.62]{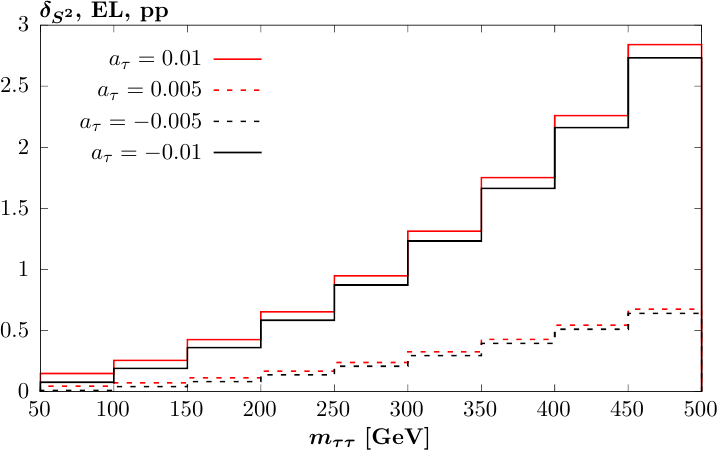}
\includegraphics[scale=0.62]{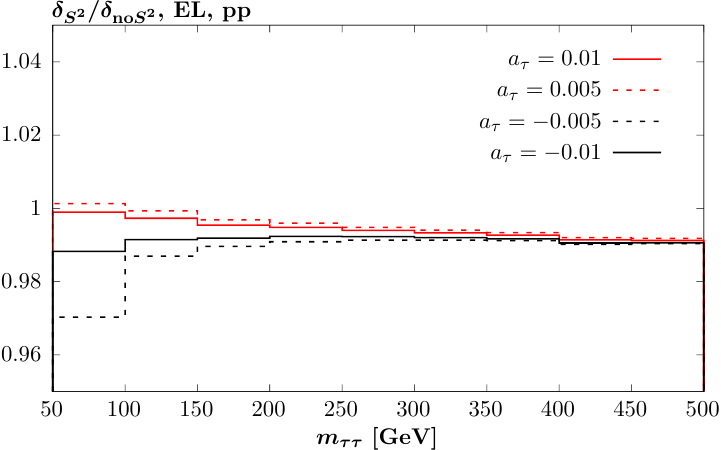}
\caption{\sf \footnotesize{(Left) As in Fig.~\ref{fig:PbPb_ai_surv} and (right) and (bottom) as in Fig.~\ref{fig:PbPb_ai_del} but now for pp collisions at 13 TeV, with cuts as described in text.}}
\label{fig:pp_ai}
\end{center}
\end{figure}

We next turn to the case of PI $\tau$ pair production in pp collisions, taking $\sqrt{s}=13$ TeV, and  $|\eta_\tau|<2.5$ and $p_\perp^\tau < 25$ GeV. In Fig.~\ref{fig:pp_ai} (left) we show the same comparison as in Fig.~\ref{fig:PbPb_ai_surv} for the PbPb case, namely the survival factor differentially, but now in the $\tau$ pair invariant mass, $m_{\tau\tau}$, for the individual components of \eqref{eq:sigtau}, again up to $O(a_\tau^2)$. While the overall size of this is different from the PbPb case, due to the differing beams and kinematics, we can see there are similarities in the  differences between the components. Namely the $O(a_\tau)$ survival factor is somewhat higher than the $O(a_\tau^0)$, while the $O(a_\tau^2)$ is close to it (and in this case somewhat below it). In Fig.~\ref{fig:pp_ai} (bottom) we show the same comparison as in Fig.~\ref{fig:PbPb_ai_del} (left), namely for the ratio of  cross section modifications, $\delta$, and as we would expect this gives similar levels of difference, at the percent level. In Fig.~\ref{fig:pp_ai} (right) we show the absolute values of the modifications, and we can see that these increase monotonically to rather large values at larger $\tau$ invariant masses. As there is no sign change in the modifications, there is in contrast to the PbPb case no enhancement in the ratios in the bottom plot. The impact of omitting this dependence on any $a_\tau$ determination will therefore likewise be at the percent level.

\begin{figure}
\begin{center}
\includegraphics[scale=0.62]{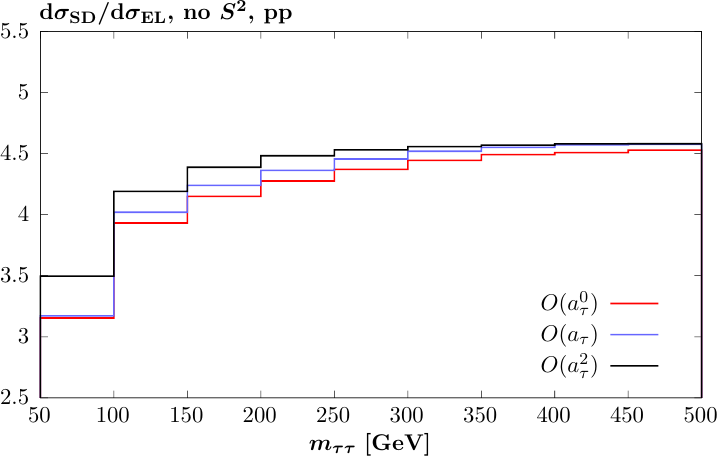}
\includegraphics[scale=0.62]{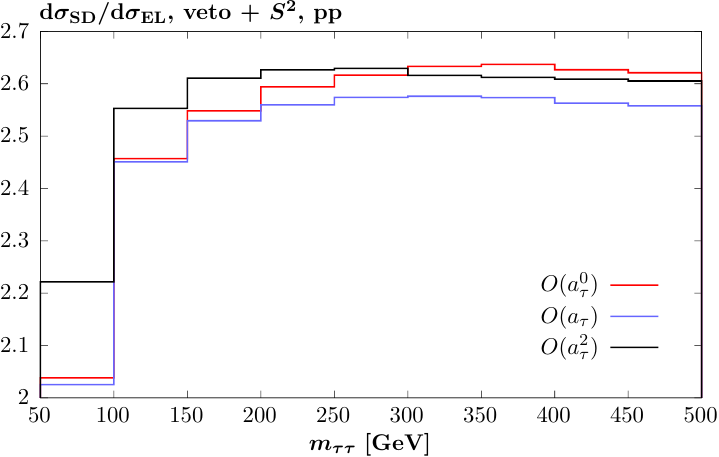}
\includegraphics[scale=0.62]{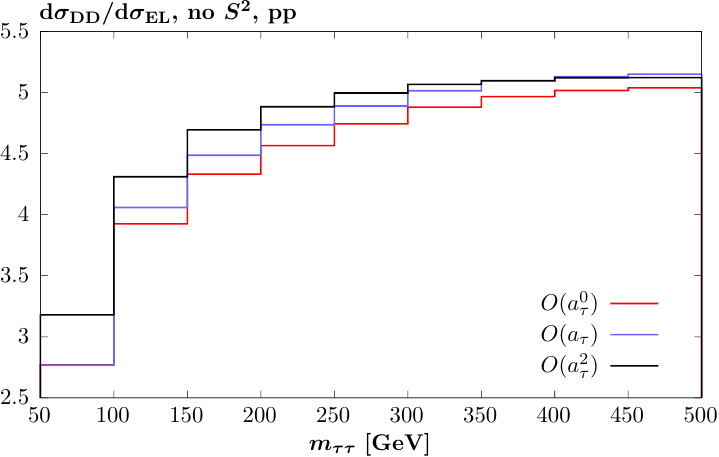}
\includegraphics[scale=0.62]{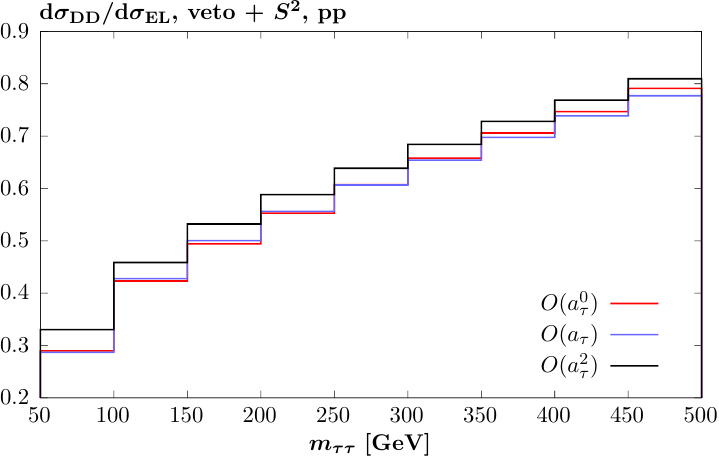}
\includegraphics[scale=0.62]{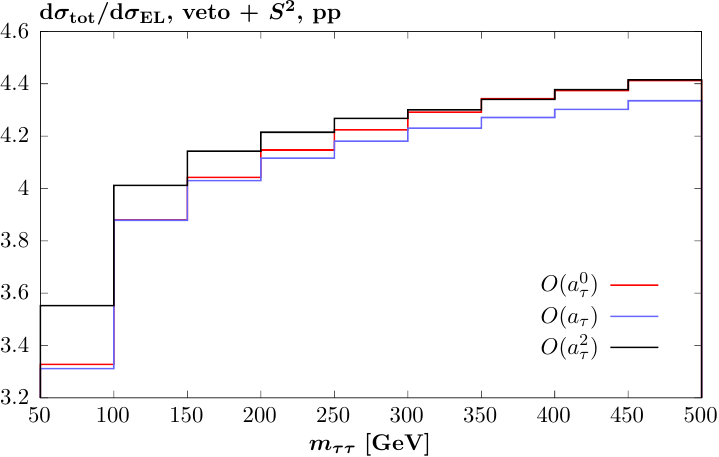}
\caption{\sf \footnotesize{Ratios of dissociative (SD, DD) and total (EL + SD + DD) to elastic (EL) $\tau$ pair production cross sections in pp collisions at $\sqrt{s}=13$ TeV, for different $O(a_\tau)$ contributions to the overall cross section. Cuts are as described in text.}}
\label{fig:pp_sddd}
\end{center}
\end{figure}

Considering the impact of proton dissociation, in Fig.~\ref{fig:pp_sddd} we show a range of ratios of the SD and DD to the EL cross sections for the different  individual components of \eqref{eq:sigtau}. These are again only shown up to $O(a_\tau^2)$, as the higher power contributions are subleading, although in this case in contrast to the survival factor the behaviour of these is distinct from  that of the $O(a_\tau^2)$ in terms of the plotted ratios. We recall that in order to select PI production, it is common to impose a veto on additional particle production in the central detector. We will account for this approximately by simply vetoing on the outgoing quark lines, assuming LO kinematics, that come on the dissociative side(s) for the SD and DD channels. We in particular veto on these for the case that $p_\perp>0.5$ GeV and $|\eta|<2.5$, though the results will not be too sensitive to this precise choice.

 In the top left and middle left plots we show the SD and DD ratios for the experimentally unrealistic case where we do not include the survival factor or any veto, for the sake of demonstration. We note that the overall size of these ratios and trend with increasing invariant mass is as we would expect broadly in line with the comparisons shown in e.g.~\cite{Harland-Lang:2020veo}, as is also the case when the veto is applied, as discussed below. In particular, prior to imposing any veto or accounting for the survival factor the dissociative contributions completely dominate over the elastic. We can also see that there is a non--negligible, up to $\sim 10\%$ level difference between the $O(a_\tau^0)$ and higher power components, with crucially the more important $O(a_\tau^2)$ being the most different from this. The difference is somewhat larger in the DD case, as we might expect given the larger average photon virtuality here. 
 
On the other hand, when a veto is imposed and the survival factor is accounted for, the DD contribution is predicted to be significantly reduced, though not  negligible in comparison to the EL at the highest masses. This is again in line with the discussion in e.g.~\cite{Harland-Lang:2020veo} and in particular the expectation that the survival factor in the DD case is significantly lower than in the EL and SD channels, due to the lower average proton impact parameter. The SD contribution is also reduced, though remains dominant, as here the survival factor is higher. We can see that broadly the impact of the veto in particular is to reduce the differences between the different  components of \eqref{eq:sigtau}, but that these are still present. 

In the bottom plot we show the prediction for the ratio of the total (EL + SD + DD) cross section. The overall size of this is broadly in line with that observed in e.g. the case of dimuon production in~\cite{CMS:2024skm}, as is the trend for this to increase with invariant mass. The differences between the different  components of \eqref{eq:sigtau} are again present, as we would expect, entering at the 10\% level. However, due to an accidental cancellation between then SD and DD components, at higher mass the $O(a_\tau^2)$ and $O(a_\tau^0)$ ratios agree rather well.

\begin{figure}
\begin{center}
\includegraphics[scale=0.62]{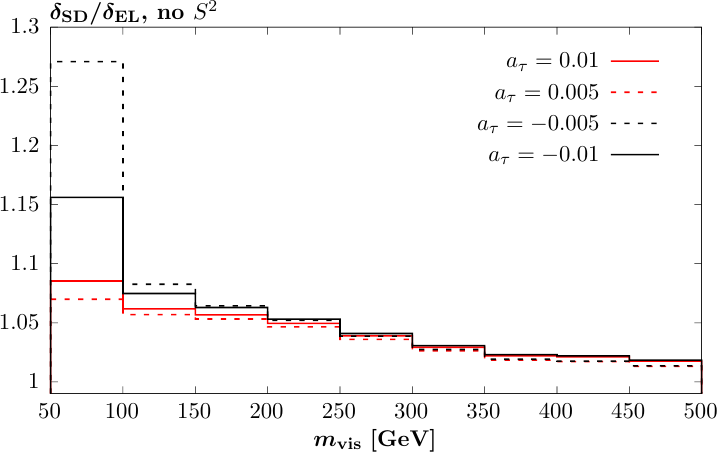}
\includegraphics[scale=0.62]{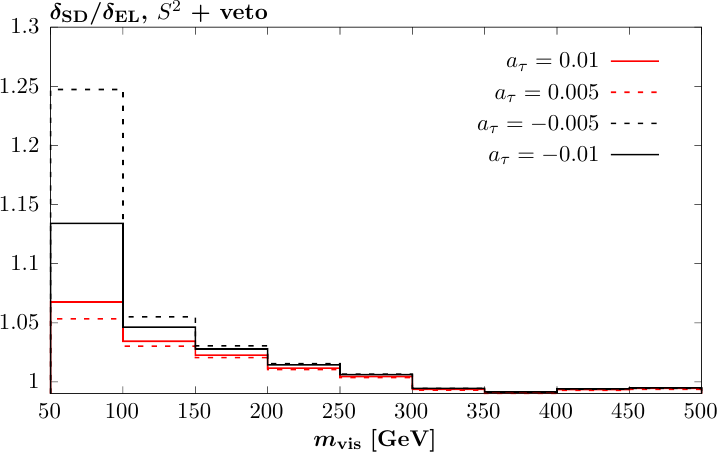}
\includegraphics[scale=0.62]{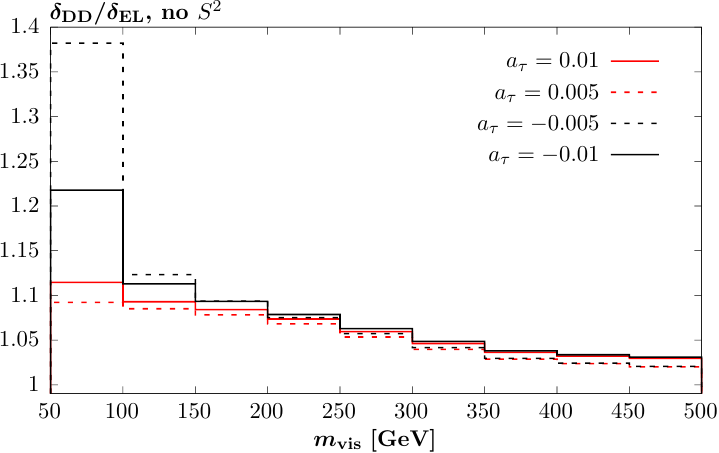}
\includegraphics[scale=0.62]{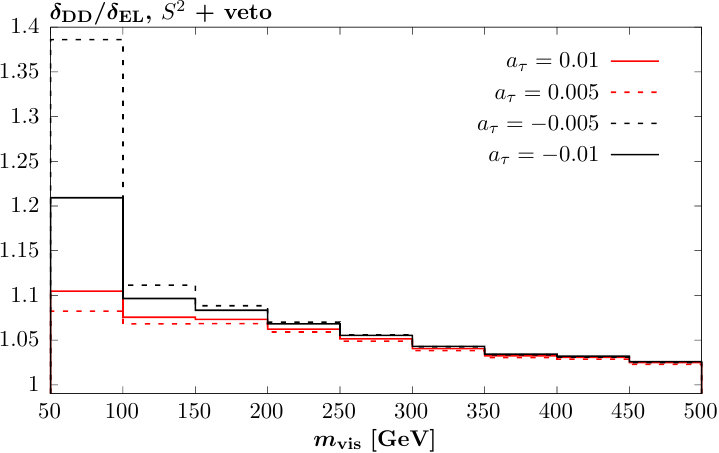}
\includegraphics[scale=0.62]{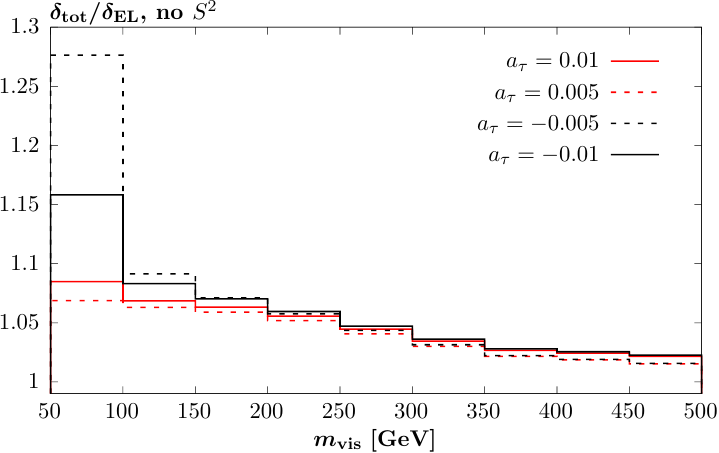}
\includegraphics[scale=0.62]{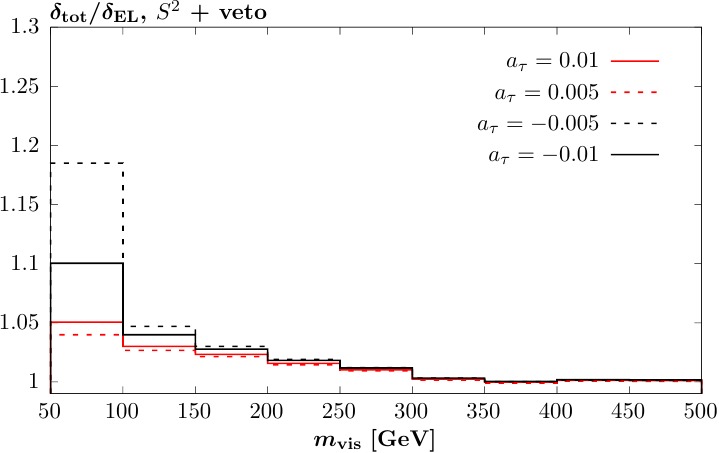}
\caption{\sf \footnotesize{Difference in cross section modifications, $\delta$, due to non--zero anomalous $a_\tau$, between the  dissociative (SD, DD) and total (EL + SD + DD)  and elastic (EL)  cases. Results shown for $\tau$ pair production in pp collisions at $\sqrt{s}=13$ TeV, as in Fig.~\ref{fig:pp_sddd}}}
\label{fig:pp_sddd_del}
\end{center}
\end{figure}

In Fig.~\ref{fig:pp_sddd_del} we show the corresponding impact on the ratio of cross section modifications, $\delta$. For the SD and DD ratios we can see $O(10\%)$ deviations at lower invariant mass, but which reduce to the percent level at higher invariant mass. This can be explained by the fact that for SD and DD production, at large $m_{\tau\tau}$ the relative impact of the larger average photon $Q_i^2$, which becomes on average much lower than $m_{\tau\tau}$, will become increasingly less significant. The impact of the veto is in line with the observations above, and tends to further reduce the deviation. 

Focussing on the experimentally relevant case of the total (EL + SD + DD) cross section ratio we can see a similar trend to the individual SD and DD cases, but once a veto is imposed the impact of the accidental cancellation noted above is clear, with the deviation in the ratio becoming almost negligible at high mass. Nonetheless, given the value of $a_\tau$ is derived from shape information that may in principle extend to the lower invariant mass values shown here the impact may not be negligible, although we note from Fig.~\ref{fig:pp_ai} (right) that the absolute size of the deviation in the lower mass bin is $O(10\%)$ and hence a 10\% change in this corresponds to a percent level change in the cross section, which is therefore again relatively mild.

\begin{figure}
\begin{center}
\includegraphics[scale=0.62]{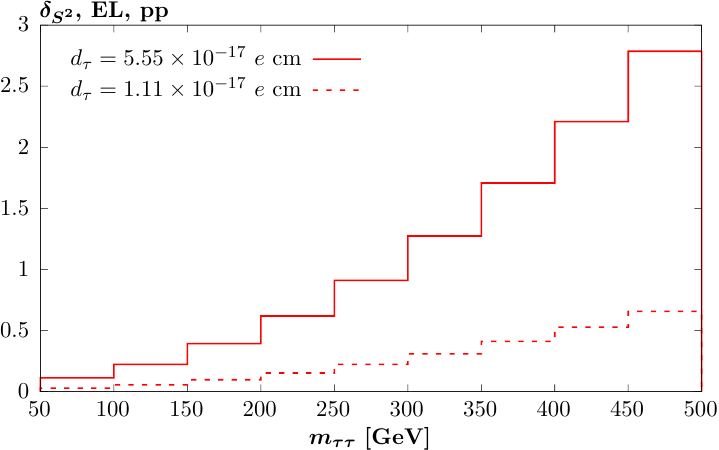}
\includegraphics[scale=0.62]{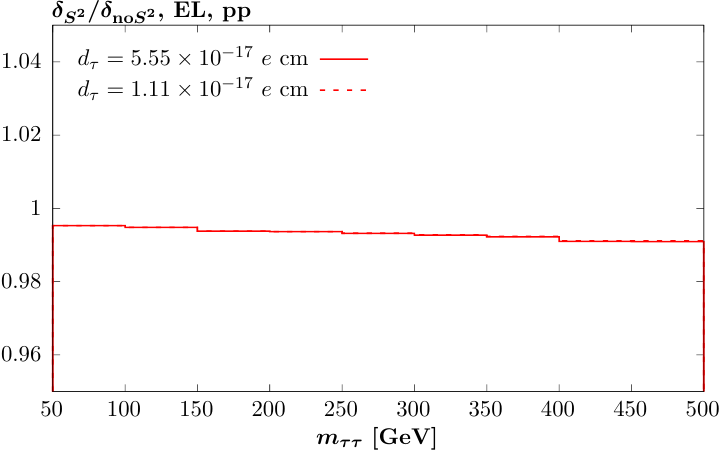}
\caption{\sf \footnotesize{As in Fig.~\ref{fig:pp_ai} (top left) and (bottom)  but now for non--zero values of $d_\tau$, chosen to match the size of the $a_\tau$ values that enter in that figure. The results for negative values are identical to the positive valued case, and hence are not shown.}}
\label{fig:pp_dtau_ai}
\end{center}
\end{figure}

\begin{figure}
\begin{center}
\includegraphics[scale=0.62]{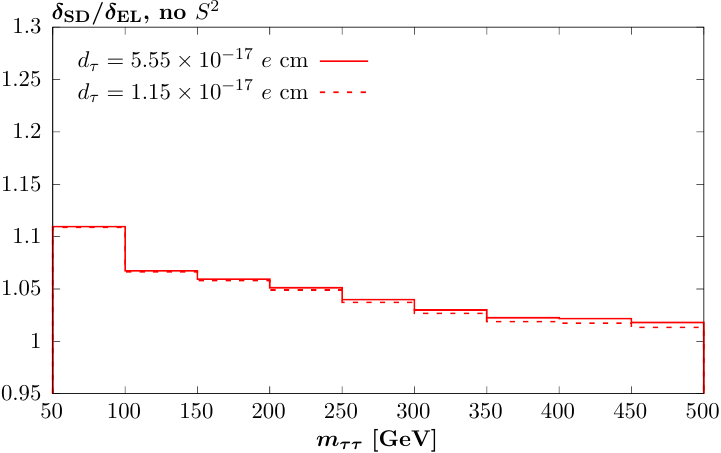}
\includegraphics[scale=0.62]{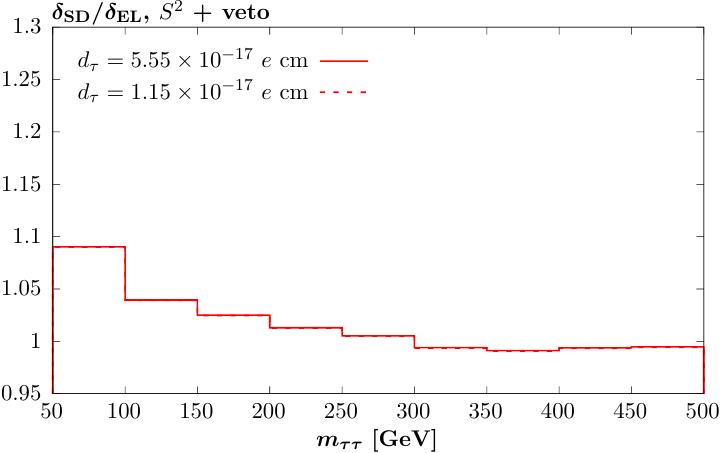}
\includegraphics[scale=0.62]{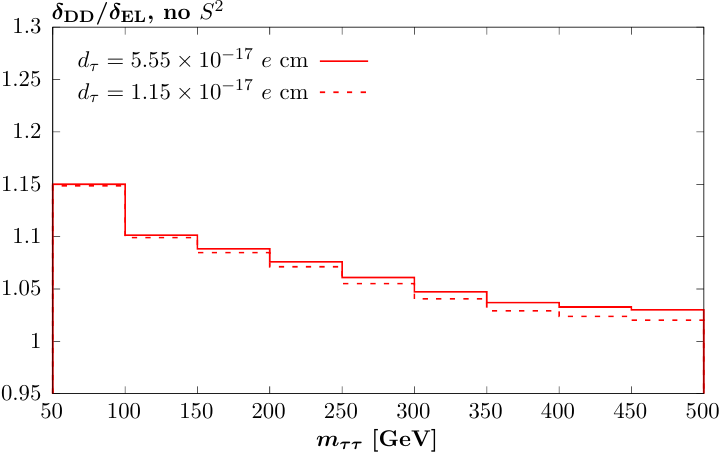}
\includegraphics[scale=0.62]{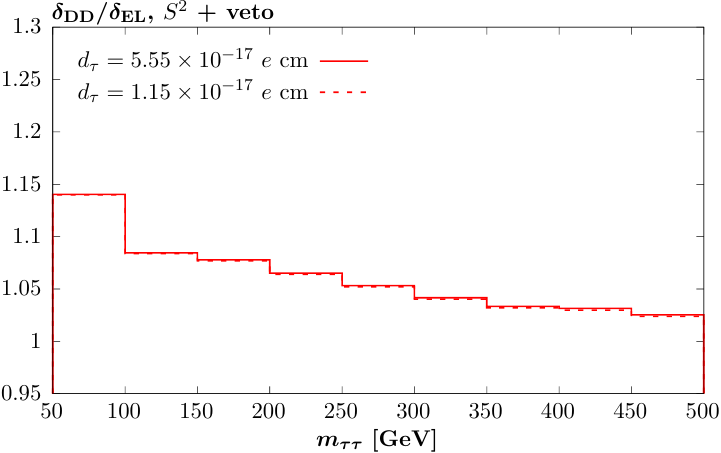}
\includegraphics[scale=0.62]{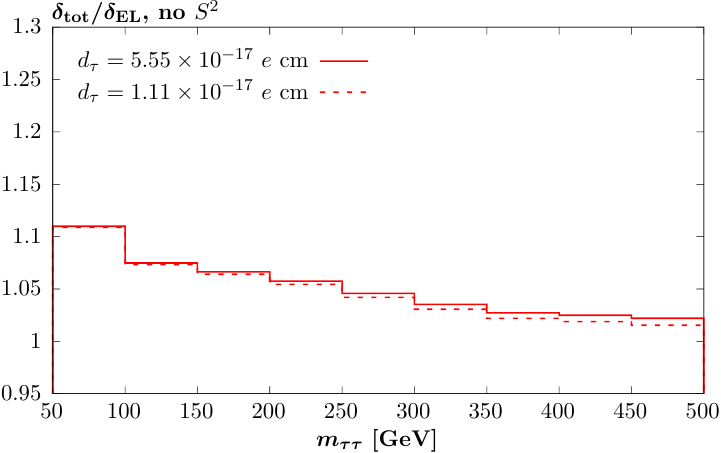}
\includegraphics[scale=0.62]{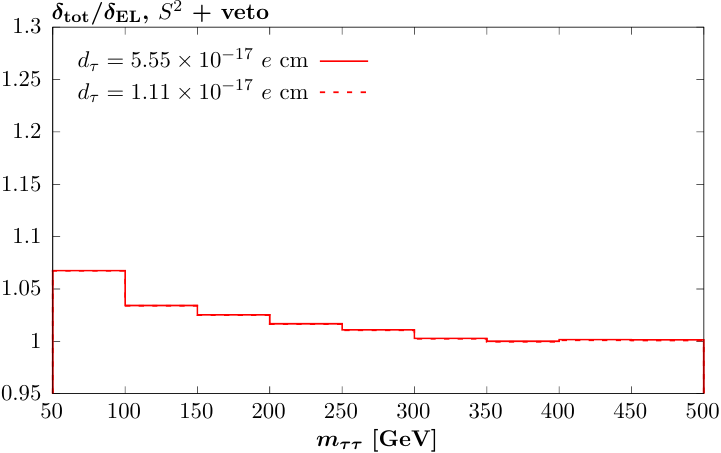}
\caption{\sf \footnotesize{As in Fig.~\ref{fig:pp_sddd_del} but now for non--zero values of $d_\tau$, chosen to match the size of the $a_\tau$ values that enter in that figure. The results for negative values are identical to the positive valued case, and hence are not shown.}}
\label{fig:pp_sddd_dtau_del}
\end{center}
\end{figure}

Finally, while the discussion above has focussed for concreteness on the case of the anomalous magnetic dipole moment, $a_\tau$, very similar observations follow for the case of the electric dipole moment. As noted above, for the case of $d_\tau$ the expansion is identical to \eqref{eq:sigtau} but the linear and cubic terms in $d_\tau$ are zero due to the presence of the $\gamma_5$ in \eqref{eq:vtau}. Given the quadratic, $O(a_\tau^2)$, terms are observed to generally dominate for the experimentally relevant values of $a_\tau$, we expect very similar trends in the case of $d_\tau$. Namely, mild deviations at the percent level in the cross section modifications for values of $d_\tau$ at the level of current experimental limits.

To confirm this, in Fig.~\ref{fig:pp_dtau_ai} we show the same comparison as in Fig.~\ref{fig:pp_ai} (top left) and (bottom)  but now for non--zero values of $d_\tau$, chosen to match the size of the $a_\tau$ values that enter in that figure. That is, such that e.g. the $O(a_\tau^2)$ and $O(d_\tau^2)$ contributions to the cross section are the same, and so on. The only difference is then due to the fact that  the odd, $O(d_\tau, d_\tau^3)$, contributions are now zero. As a result, the contribution for positive and negative values of $d_\tau$ is identical, and so only the positive case is now shown. We can see that both the absolute impact of the cross section modifications and the difference that comes from a complete treatment of the survival factor is very similar to the case of non--zero $a_\tau$. This is as expected, given the $O(a_\tau^2)$ contribution to the modification is dominant and is by construction the same between the $O(a_\tau^2)$  and $O(d_\tau^2)$ cases.

We next turn to the case of proton dissociation, showing in Fig.~\ref{fig:pp_sddd_dtau_del} the same plot as Fig.~\ref{fig:pp_sddd_del} but again for non--zero values of $d_\tau$, chosen to match the size of the $a_\tau$ values that enter in that figure. The difference with respect to the non--zero $a_\tau$ case is somewhat larger than in the previous comparison, due to the larger role of the odd, $O(a_\tau)$, contribution in particular. Nonetheless, qualitatively we can see that a very similar picture emerges.

 \section{Summary and Outlook}\label{sec:conc}
 
 In this paper we have presented the   the first calculation of photon--initiated $\tau$ pair production in the presence of non--zero anomalous magnetic ($a_\tau$) and or electric dipole ($d_\tau$) moments of the $\tau$ lepton that accounts for the non--trivial interplay between these modifications with the soft survival factor and the possibility of dissociation of the hadron (proton or ion) beam. The impact of both of these effects is on general grounds not expected to have a uniform dependence on the value of $a_\tau, d_\tau$, but in all previous analyses this assumption has been made. 
 
 We have therefore investigated the importance of these effects in the context of photon--initiated $\tau$ pair production in both pp and PbPb collisions. This is in general found to be small, at the percent level in terms of any extracted limits or observations of $a_\tau, d_\tau$, such that these effects can indeed be safely ignored in existing experimental analyses. However, as the precision of such determinations increases in the future, the relevance of these effects will likewise increase. With this in mind we have made our calculation publicly available in the \texttt{SuperChic} Monte Carlo (MC) generator:
\begin{center}
    \href{https://github.com/LucianHL/SuperChic}{https://github.com/LucianHL/SuperChic}.
\end{center}
This simulates photon--initiated $\tau$ pair production for arbitrary values of $a_\tau, d_\tau$ in both pp and heavy ion collisions. In the former case both elastic and inelastic (single and double dissociative) production are fully modelled, including the dependence of these different contribution on the  $a_\tau, d_\tau$. In the latter case mutual ion dissociation is also accounted for. In both cases the impact of the survival factor and the interplay with  $a_\tau, d_\tau$ is taken into account. In addition, the $a_\tau, d_\tau$ dependence of the predicted cross section can be provided in a manner such that the result for different values of $a_\tau, d_\tau$ can be automatically provided without the need for additional running of the MC. We note that the interface settings to \texttt{Pythia} for PI production are also updated in this latest version; more details can be found in the user manual provided in the above repository.

The prospects for future limit setting and determinations of the anomalous magnetic and or electric dipole moments of the $\tau$ lepton in the photon--initiated channel, and for physics studies in this channel more generally, are very promising. As such, it is important that the theoretical framework underpinning these experimental analyses is as precise and robust as possible. The work presented here and in the updated version of \texttt{SuperChic} have provided a key element in that effort.

\section*{Acknowledgements}
I am grateful to Valery Khoze and Misha Ryskin for comments on the manuscript and useful discussions and to Ilkka Helenius for assistance with Pythia parton shower settings. I thank  the Science and Technology Facilities Council (STFC) part of U.K. Research and Innovation for support via the grant award ST/T000856/1.

\bibliography{references}{}
\bibliographystyle{h-physrev}

\end{document}